\documentclass[aps,eqsecnum,preprintnumbers,preprint]{revtex4}
\usepackage{axodraw}
\newcommand{\beq}{\begin{equation}}
\newcommand{\eeq}{\end{equation}}
\newcommand{\beqa}{\begin{eqnarray}}
\newcommand{\eeqa}{\end{eqnarray}}
\newcommand{\lslash}[1]{#1\llap/}

\newcommand{\Tr}{\mathop{\rm Tr}}

\newcommand{\Eq}[1]{Eq.\ (\ref{#1})}
\newcommand{\Eqs}[2]{Eqs.\ (\ref{#1}) and (\ref{#2})}
\newcommand{\Eqss}[3]{Eqs.\ (\ref{#1}),  (\ref{#2}) and (\ref{#3})}
\newcommand{\Ref}[1]{Ref.\ \cite{#1}}
\newcommand{\Fig}[1]{Fig.\ \ref{#1}}

\newcommand{\pieff}{\hat\pi}

\begin{document}


\title{Thermal Field Theory in a layer: Applications of
Thermal Field Theory methods to the propagation
of photons in a two-dimensional electron sheet}

\author{Jos\'e F. Nieves}
\affiliation{Laboratory of Theoretical Physics,
Department of Physics, P.O. Box 23343,
University of Puerto Rico,
R\'{\i}o Piedras, Puerto Rico 00931-3343}

%
%
\begin{abstract}

We apply the Thermal Field Theory methods to study
the propagation of photons in a plasma layer, that is a plasma
in which the electrons are confined to a two-dimensional plane sheet.
We calculate the photon self-energy
and determine the appropriate expression for the photon propagator
in such a medium, from which the properties of the propagating
modes are obtained. The formulas for the photon dispersion relations
and polarization vectors are derived explicitly in some detail for some
simple cases of the thermal distributions of the charged particle gas,
and appropriate formulas that are applicable in more general situations
are also given.

\end{abstract}
\maketitle
%
%
\section{Introduction}
\label{sec:introduction}

It is well known that when elementary particles,
such as photons or neutrinos, propagate through a medium,
the effects of the background particles influence their properties
in important ways.
Among the various approaches that exist to study the effects of a medium
on the properties and interactions of particles, the methods of
\emph{Thermal Field Theory} (TFT) have proven to be very useful.
Largely motivated by the original work of
Weldon\cite{weldon:cov,weldon:fermions,weldon:img}, these methods
have been applied to the problems mentioned above and other similar ones,
and they have been helpful for understanding the physics involved and also
from a computational point of view.

The present work is concerned with similar calculations,
but with the distinction that the medium consists
a gas of particles that are confined to live in a
two-dimensional plane sheet, or layer. A typical system of this type
is an ordinary plasma in which the electrons are confined to a plane
sheet, that we can take to be the $z = 0$ plane. The quantities of
interest are the usual ones such as, for example, the dispersion
relations of the propagating photon modes, the damping
and the transition rates for various processes.

Before continuing we want to stress that this is not the same thing as what
is usually called $QED_3$ (or $QED$ in 2+1 dimensions), which has
been studied in the literature\cite{KleinKreisler:1994wr,Hott:1998qu}.
$QED_3$ describes a system that, with regard to the space coordinates,
has cylindrical symmetry and the physics, being independent of the
$z$ coordinate, can be studied by considering a two-dimensional cross section.
Thus, for example, the \emph{electron} in $QED_3$ is really
a line of charge in the three-dimensional world, and the Coulomb potential
between two such \emph{electrons} is logarithmic. In contrast, in
the system we are considering, the electron is an ordinary point charge,
which is confined to the $z = 0$ plane, but the Coulomb potential between
two electrons is the usual $1/r$ potential.

The method that we apply here to study these systems could be useful
in the context of astrophysical\cite{lyutikov,asseo,gedalin}
as well as plasma physics\cite{jiang} and condensed matter\cite{nagao,uchida}
applications, and they are also interesting in their own right because
they can be useful in the study of physical systems of current interest
in which a plasma is confined to a layer\cite{caldas} or a wire\cite{faccioli}.

In the context of TFT, the distinctive feature of the system
that we are considering is that the medium is not isotropic over
the (three-dimensional) space. Consequently, the thermal propagators
that are used ordinarily in TFT calculations for the case of homogenous
and isotropic media, are not the appropriate ones for the present case.
Therefore, in order to use the TFT methods to study the model of
the two-dimensional plasma layer, a crucial requirement is finding
the appropriate set of thermal propagators that must be used.

In the present work we considered the simplest situation of an ordinary
gas of electrons, which are confined to a plane sheet, but are otherwise
free. Our main goal has been to formulate the TFT
approach to the model of the two-dimensional plasma layer that we
have described. The important steps to this end are taken
in the first part of the present paper, where we determine
the appropriate set of thermal propagators. The charged particle
propagators are very similar in form to the standard three-dimensional form.
But, as we will see, the propagation of the photon in the layer
is described by en \emph{effective} field which has a corresponding propagator
that is very different from the usual one. The photon propagator is an
important quantity because its inverse determines the bilinear part
of the effective action or, equivalently, the equation of motion for the
photon effective field, from which the dispersion relations and wave
functions of the propagating photon modes can be obtained.
As an application, in the second part we carry out a one-loop
calculation of the photon self-energy in that medium and,
as a specific example, we consider in detail the calculation
of the longitudinal dispersion relation. There we compare our approach
and results for this calculation with the results that are known for this
system in literature\cite{bret,fetter}, which
have been obtained using the so-called
\emph{static local field correction approximation}\cite{ichimaru2}.
There we show explicitly that
our results for this calculation reduce to those known results
when the appropriate limits are taken and/or approximations are made,
which in particular involves approximating the photon propagator
by its static (zero frequency) limit.
However, the formulas that we obtain, for the self-energy and dispersion
relations in general and for the longitudinal dispersion relation
in particular, can be used for a wider range of conditions in which those
approximations (such as the \emph{static local field correction approximation})
and limits are not justified.
%
%
\section{Notation and kinematics}
\label{sec:kinematics}

As usual, we denote by $u^\mu$ the velocity four-vector of the medium.
Adopting the frame in which the medium is at rest, we set
\beq
u^\mu = (1, \vec 0) \,,
\eeq
and from now on all the vectors refer to that frame. Thus,
denoting by $\vec n$ the unit vector perpendicular to the plane layer,
we introduce the four-vector
\beq
n^\mu = (0, \vec n) \,,
\eeq
and denote the momentum four-vector of a photon that propagates in the plane is
by
\beq
\label{kperpdef}
k^\mu_\perp = (\omega, \vec \kappa_\perp) \,,
\eeq
where
\beq
\vec \kappa_\perp\cdot \vec n = 0\,.
\eeq
It is useful to define
\beq
\tilde u_{\mu} \equiv u_\mu - \frac{(u\cdot k_\perp) k_{\perp\mu}}{k^2_\perp}\,,
\eeq
as well as the tensors
\beqa
\label{gperpQTdef}
\label{gtilde}
\tilde g_{\perp\mu\nu} & = & g_{\mu\nu} -
	\frac{k_{\perp\mu} k_{\perp\nu}}{k^2_\perp} +
	n_\mu n_\nu \,,\nonumber\\
Q_{\mu\nu} & = & \frac{\tilde u_\mu \tilde u_\nu}{\tilde u^2} \,,\nonumber\\
T_{\mu\nu} & = & \tilde g_{\perp\mu\nu} - Q_{\mu\nu} \,.
\eeqa
It is useful to note that the tensors $Q$ and $T$ are symmetric
and satisfy
\beqa
k^\mu_\perp T_{\mu\nu} = n^\mu T_{\mu\nu} = \tilde u^\mu T_{\mu\nu} 
	& = & 0\,,\nonumber\\
k^\mu_\perp Q_{\mu\nu} = n^\mu Q_{\mu\nu} & = & 0\,,
\eeqa
as well as
\beqa
Q_{\mu\lambda} {Q^\lambda}_\nu & = & Q_{\mu\nu} \,,\nonumber\\
T_{\mu\lambda} {T^\lambda}_\nu & = & T_{\mu\nu} \,,\nonumber\\
T_{\mu\lambda} {Q^\lambda}_\nu & = & 0\,.
\eeqa
%
%
%
\section{The Model}

\subsection{The electron free field}

We envisage a slab of area $L^2$ in the $xy$ plane
and thickness $L_z$. The electrons are confined to live within that slab
but are otherwise free to move within it. Eventually, the limits
$L \rightarrow \infty$ and $L_z \rightarrow 0$ will be taken in a suitable
way.

Making use of the Furry picture, the electron field is
expanded in terms of the one-particle wavefunctions which we take
to be of the form
\beq
H(z)e^{i\vec p_\perp\cdot \vec x_\perp} \,,
\eeq
where $\vec p_\perp$ is the component of $\vec p$ in the xy plane
and similarly for $\vec x_\perp$. The wavefunctions can be taken to
satisfy periodic boundary conditions in the $xy$ plane as usual.
In principle the function $H(z)$ is
labeled by some quantum number. However, since we are contemplating
taking the $L_z \rightarrow 0$ limit, for our purposes we assume 
that only the lowest lying state survives. Moreover, for the same reason,
the particular form that it may take is not relevant, as long as it is
consistent with our assumptions, and in particular we can adopt
\beq
H(z) = \left\{
\begin{array}{ll}
1 & \mbox{for} -\frac{L_z}{2} \le z \le \frac{L_z}{2} \\[12pt]
0 & \mbox{otherwise}
\end{array}
\right.
\eeq

The main assumption of the model is that in the passage to the continuous
momentum plane waves in the $xy$ plane ($L\rightarrow \infty$),
the current density becomes
\beq
\bar\psi\gamma_\mu\psi = \left(\frac{H(z)}{\sqrt{L_z}}\right)^2
\bar{\hat\psi}\gamma_{\perp\mu}\hat\psi\,,
\eeq
where the $\gamma^\mu_\perp$ are the gamma matrices that are perpendicular
to $n^\mu$, i.e.,
\beq
\gamma_\perp \cdot n = 0\,,
\eeq
with the electron field of the form
\beqa
\label{psiperpexp}
\hat\psi(x_\perp) & = & \int\frac{d^2p_\perp}{(2\pi)^2 2E}\left[
a(\vec p_\perp,s) u(\vec p_\perp,s) e^{-ip_\perp\cdot x_\perp} +
b^\ast(\vec p_\perp,s) v(\vec p_\perp,s) e^{ip_\perp\cdot x_\perp}\right]\,.
\eeqa
In \Eq{psiperpexp},
$E = \sqrt{\vec p^{\,2}_\perp + m^2}$, $x^\mu_\perp = (x^0,\vec x_\perp)$
and $p^\mu_\perp = (E,\vec p_\perp)$. The spinors $u$ are
the standard Dirac spinors normalized such that
\beq
u\bar u = 2m\,,
\eeq
and the creation and annihilation operators satisfy
\beq
\left\{a(\vec p_\perp,s),a^\ast(\vec p^{\,\prime}_\perp,s^\prime)\right\} = 
(2\pi)^2 2E\, \delta^{(2)}(\vec p_\perp - \vec p^{\,\prime}_\perp)
\delta_{s,s^\prime} \,,
\eeq
with analogous relations for the spinors $v$ and the $b$ operators.

The passage to the planar layer is  made by taking the
$L_z \rightarrow 0$ limit at the appropriate stage, which
in essence reduces to use
\beqa
\label{Hlimit}
\frac{H(z)}{L_z} & \rightarrow & \delta(z) \nonumber\\
H(0) & = & 1 \,,
\eeqa
where $\delta(z)$ is the Dirac delta function, and the free-field current
density operator then takes the form
\beq
\label{freefieldj}
\bar\psi\gamma_\mu\psi = \delta(z)\bar{\hat\psi}\gamma_{\perp\mu}\hat\psi\,,
\eeq
For practical purposes Eqs. (\ref{psiperpexp}) and (\ref{freefieldj})
can be taken as the equations that define our model for the plane layer.

\subsection{Electron thermal propagator}
\label{sec:electronpropagator}

From the point of view of TFT, the important ingredient for carrying
out the calculation of the photon self-energy is the
electron thermal propagator, which in turn can be determined by
well established rules in terms of the free-field propagator.
The propagator associated with the electron free-field  $\psi_\perp$ can
be determined in various ways, in analogy with the usual case,
and the results look similar. The various components of the
thermal propagator matrix are simply
\beqa
\label{fermionpropagator}
S_{\perp 11}(p_\perp) & = & (\lslash{p}_\perp + m_e)\left[
\frac{1}{p^2_\perp - m^2_e + i\epsilon} +
2\pi i\delta(p^2_\perp - m^2_e)\eta_e(p_\perp\cdot u)\right] \,,\nonumber\\
S_{\perp 22}(p_\perp) & = & (\lslash{p}_\perp + m_e)\left[
\frac{-1}{p^2_\perp - m^2_e - i\epsilon} +
2\pi i\delta(p^2_\perp - m^2_e)\eta_e(p_\perp\cdot u)\right] \,,\nonumber\\
S_{\perp 12}(p_\perp) & = & (\lslash{p}_\perp + m_e)2\pi i\left[
\eta_e(p_\perp\cdot u) - \theta(-p_\perp\cdot u)\right]\,,\nonumber\\
S_{\perp 21}(p_\perp) & = & (\lslash{p}_\perp + m_e)2\pi i\left[
\eta_e(p_\perp\cdot u) - \theta(p_\perp\cdot u)\right]\,,
\eeqa
where
\beq 
\label{etae} 
\eta_e(p) = \theta(p\cdot u)f_e(p\cdot u) +
\theta(-p\cdot u)f_{\bar e}(-p\cdot u)\,, 
\eeq
with
\beqa 
f_e(x) & = & \frac{1}{e^{\beta(x - \mu_e)} + 1} \nonumber\\
f_{\bar e}(x) & = & \frac{1}{e^{\beta_e(x + \mu_e)} + 1}
\eeqa
and $\theta(x)$ is the step function.
Here $\beta_e$ and $\mu_e$ are the inverse temperature and the
chemical potential of the electron gas, respectively.
The total number of particles in the gas can be calculated from
\beqa
N & = & \int d^3x\,\bar\psi\gamma^0 \psi \nonumber\\
& = & L^2 \int\frac{d^3p_\perp}{(2\pi)^3}
\mbox{Tr}\left[S_{\perp 11}(p_\perp)\gamma^0\right]\,,
\eeqa
where we have set $\int dx\,dy \rightarrow L^2$.
Using the formulas given above for the propagator, this yields
\beq
N/L^2 = n_e + n_{\bar e} \,,
\eeq
where
\beq
n_{e,\bar e} = 2\int\frac{d^2\vec p_\perp}{(2\pi)^2}
\frac{1}{e^{\beta(E \mp \mu_e)} + 1}\,,
\eeq
which represent the surface density of electrons and positrons respectively.
%
%
\section{Photon propagation in the layer}

\subsection{Photon effective field}
\label{sec:photoneffectivefield}

Using \Eq{freefieldj} as the starting point, the interaction Lagrangian term
in the action is taken to be
\beq
\label{Sint}
S_{\mbox{int}} = - e\int d^3x_\perp\;\bar{\hat\psi}\gamma^\mu_\perp\hat\psi
\hat A_\mu \,,
\eeq
where
\beq
\hat A_\mu \equiv \left. A_{\perp\mu} \right|_{z = 0} \,.
\eeq
This indicates that $\hat A_\mu$
is the relevant electromagnetic field variable, and it is the one that we
should focus on. Thus, we regard $\hat A_\mu$ as the effective field
for the photon, and our goal is to determine its effective action,
or equivalently its equation of motion, including the thermal corrections.

Formally, what we want to do is to integrate out all the dynamical field
variables except $\hat A_\mu$ itself. Following the usual
functional method of quantization of the electromagnetic field\cite{peskin},
and adapting it to the present model, a convenient way to proceed is to
introduce in the action an external current
\beq
J^\mu(x) = \delta(z) \hat J^\mu(x_\perp) \,,
\eeq
with $\hat J^\mu(x_\perp)$ satisfying
\beqa
\label{Jtranscond}
\hat J(x_\perp) \cdot n & = 0\,, \nonumber\\
\partial_\perp\cdot \hat J(x_\perp) & = & 0\,,
\eeqa
where $\partial^\mu_\perp = (\frac{\partial}{\partial x_0},
-\frac{\partial}{\partial\vec x_\perp})$.
\Eq{Jtranscond} ensures that we are selecting only the transverse
(gauge invariant) part of $\hat A_\mu$, which is the physically meaningful
one, since the longitudinal part decouples.
The \emph{classical} field, which we denote by $A^{(J)}_\mu$,
in the presence of both, the external current $\hat J_\mu$ and the interaction
given by $S_{\mbox{int}}$ in \Eq{Sint}, is then defined by
\beq
A^{(J)}_\mu = \frac{1}{Z}\frac{i\delta Z}{\delta \hat J^\mu} \,.
\eeq

\subsection{Photon propagator}
\label{sec:photonpropagator}

Following the usual argument, and remembering \Eq{Hlimit},
the generating functional for the
free electromagnetic field in the layer is
\beq
Z[\hat J] \propto \exp\left\{-\frac{i}{2}\int d^3x_\perp d^3x^\prime_\perp
\hat J^\mu(x_\perp)\hat\Delta_{F\mu\nu}(x_\perp - x^\prime_\perp)
\hat J^\nu(x^\prime_\perp)
\right\}\,,
\eeq
where $\hat\Delta_{F\mu\nu}(x_\perp - x^\prime_\perp)$ is obtained from the
standard photon propagator $\Delta_{F\mu\nu}(x - x^\prime)$
by setting the coordinates normal to the plane
($z$ and $z^\prime$) equal to zero. Therefore,
taking into account \Eq{Jtranscond},
the propagator for the effective field in the plane is given,
in momentum space, by
\beq
\hat\Delta_{F\mu\nu}(k_\perp) = \tilde g_{\perp\mu\alpha}
\tilde g_{\perp\nu\beta}\left(
\int\frac{d\kappa_\parallel}{2\pi}\; \Delta_F^{\alpha\beta}(k)\right) \,,
\eeq
where $\tilde g_{\perp\mu\nu}$ has been defined in \Eq{gtilde} and,
in the integrand, the momentum vector $k$ is decomposed in the form
\beq
k_\mu = k_{\perp\mu} + \kappa_\parallel n_\mu\,,
\eeq
with $k_{\perp\mu}$ as given in \Eq{kperpdef}. Writing
\beq
\Delta_{F\mu\nu}(k) = \frac{-g_{\mu\nu}}{k^2 + i\epsilon} + 
\mbox{gauge-dependent terms} \,,
\eeq
and carrying out the integration over $\kappa_\parallel$,
we then obtain the propagator for the effective photon field as
\beq
\label{effectivefreeprop}
\hat \Delta_{F\mu\nu}(k_\perp) = -\hat\Delta(k_\perp)\left(
T_{\mu\nu} + Q_{\mu\nu}\right)\,,
\eeq
where the tensors $T$ and $Q$ have been defined in \Eq{gperpQTdef}, and
\beq
\label{effectivephotonpropagator}
\hat\Delta(k_\perp) = \left\{
\begin{array}{ll}
\frac{-i}{2\sqrt{\omega^2 - \kappa^2}} & \mbox{for $\omega > \kappa$} \\[12pt]
\frac{-1}{2\sqrt{\kappa^2 - \omega^2}} & \mbox{for $\omega < \kappa$} \,,
\end{array}
\right.
\eeq
with
\beq
\kappa = |\vec \kappa_\perp| \,.
\eeq

\subsection{Equation of motion}
\label{sec:eqofmotion}

Armed with the expression for the \emph{free} photon propagator
in the plane, the bilinear part of the effective action for $A^{(J)}_\mu$
is then given, in momentum space, by
\beq
S^{(2)} = \int \frac{d^3 k_\perp}{(2\pi)^3}\left\{
\frac{1}{2}A^{(J)\ast}_\mu(k_\perp) \left[D^{\mu\nu}(k_\perp) +
\hat\pi^{\mu\nu}(k_\perp)\right] A^{(J)}_\nu(k_\perp) -
A^{(J)\ast}(k_\perp)\cdot \hat J(k_\perp)\right\} \,,
\eeq
where $D^{\mu\nu}(k_\perp)$ is defined
\beq
\hat \Delta_F^{\mu\lambda}(k_\perp) D_{\lambda\nu}(k_\perp) = 
\tilde g^\mu_{\perp\nu} \,,
\eeq
and $\pieff_{\mu\nu}$ is the photon self-energy in the medium.
The dispersion relations, and the corresponding polarization vectors,
of the propagating photon modes can be determined by solving the equation
of motion for the \emph{classical} field in the absence of the external
current, that is
\beq
\label{eqofmotion}
\left[D^{\mu\nu}(k_\perp) +
\hat\pi^{\mu\nu}(k_\perp)\right] A^{(0)}_\nu(k_\perp) = 0\,.
\eeq

\subsection{Photon self-energy and dispersion relations}
\label{sec:photonselfenergy}

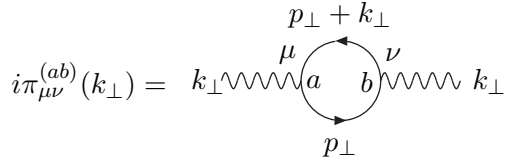
\begin{figure}
\begin{center}
\begin{picture}(160,120)(-50,0)
\Photon(30,60)(0,60){3}{5}
\Photon(90,60)(60,60){3}{5}
\LongArrowArc(45,60)(15,-93,98)
\LongArrowArc(45,60)(15,98,278)
\Text(45,85)[]{$p_\perp + k_\perp$}
\Text(45,35)[]{$p_\perp$}
\Text(-50,60)[]{$i\pi^{(ab)}_{\mu\nu}(k_\perp) =$}
\Text(65,70)[]{$\nu$}
\Text(25,70)[]{$\mu$}
\Text(35,60)[]{$a$}
\Text(55,60)[]{$b$}
\Text(-5,60)[]{$k_\perp$}
\Text(95,60)[l]{$k_\perp$}
\end{picture}
\end{center}
\caption[]{One-loop diagram for the photon thermal self-energy matrix.
\label{fig:oneloopdiagram}
}
\end{figure}

We denote the components of the thermal
self-energy matrix by $\pi^{(ab)}_{\mu\nu}$,
which to the lowest order are determined
by calculating the one-loop diagram shown in \Fig{fig:oneloopdiagram}.
The physical self-energy function that appears in \Eq{eqofmotion}
is determined from the
relations\footnote{See, for example, \Ref{dolivo-nieves:nucpi}
and references therein.}
\beqa
\label{pieff}
\mbox{Re}\,\pieff_{\mu\nu}(k_\perp) & = &
\mbox{Re}\,\pi^{(11)}_{\mu\nu}(k_\perp)\nonumber\\
\mbox{Im}\,\pieff_{\mu\nu}(k_\perp) & = &
\frac{i\pi^{(12)}_{\mu\nu}(k_\perp)}{2n_\gamma}\,,
\eeqa
where
\beq
n_\gamma = \frac{1}{e^{\beta k_\perp\cdot u} - 1} \,.
\eeq

In general $\pieff_{\mu\nu}$ satisfies the transversality condition
\beq
\label{transversalitycond}
k^\mu_\perp\pieff_{\mu\nu} = k^\nu_\perp\pieff_{\mu\nu} = 0\,,
\eeq
as a consequence of the conservation of the electromagnetic current and,
as we will see in Section\ \ref{sec:oneloop},
in the one-loop approximation $\pieff_{\mu\nu}$ satisfies in addition
\beq
\label{transversalitycond2}
n^\mu\pieff_{\mu\nu} = n^\nu\pieff_{\mu\nu} = 0\,.
\eeq
It then follows that, in the one loop approximation,
$\pieff_{\mu\nu}$ can be expressed in the form\cite{weldon:cov}
\beq
\label{pidecomp}
\pieff_{\mu\nu} = \pi_T T_{\mu\nu} + \pi_L Q_{\mu\nu} \,.
\eeq
In the more general case, $\pieff_{\mu\nu}$ can contain more terms such
as\cite{pisubpi,pisubpiE}
\beq
P_{\mu\nu} \equiv \frac{i}{\kappa}\epsilon_{\mu\nu\alpha\beta}
u^\alpha k^\beta_\perp\,,
\eeq
and others, but since we are considering only the one-loop
approximation to $\pieff_{\mu\nu}$,
for our purposes \Eq{pidecomp} is the
most general form.

Using \Eq{effectivefreeprop}, $D^{\mu\nu}(k_\perp)$ is given by
\beq
D^{\mu\nu}(k_\perp) = -\hat\Delta^{-1}(k_\perp)\left(
T_{\mu\nu} + Q_{\mu\nu}\right)\,,
\eeq
and, remembering \Eq{pidecomp}, the dispersion relations
\beqa
\label{longdisprel}
\hat\Delta^{-1}(k_\perp) - \pi_{L}(k_\perp) & = & 0 \,,\\
\label{transdisprel}
\hat\Delta^{-1}(k_\perp) - \pi_{T}(k_\perp) & = & 0
\eeqa
follow, where $\Delta(k_\perp)$ is given in \Eq{effectivephotonpropagator}.
%
%
\section{One-loop calculation}
\label{sec:oneloop}

\subsection{One-loop formula for the self-energy}
\label{sec:oneloopselfenergy}

We want to apply the results of the one-loop calculation of the
photon self-energy, to the dispersion relations given in
\Eqs{longdisprel}{transdisprel}. We will restrict ourselves
to the real part of the dispersion relations, and therefore
we consider here the calculation of $\pi^{(11)}_{\mu\nu}$,
from which the real part of the physical self-energy
is determined by means of \Eq{pieff}.

Referring to Fig. \ref{fig:oneloopdiagram},
\beq
\label{pi11}
i\pi^{(11)}_{\mu\nu}(k_\perp) = (-1)(-i)^2 \Tr\int\frac{d^3p_\perp}{(2\pi)^3}
\gamma_{\perp\mu} iS_{\perp11}(p_\perp + k_\perp)
\gamma_{\perp\nu} iS_{\perp11}(p_\perp) \,.
\eeq
When the formula for $S_{\perp11}$ given in
\Eq{fermionpropagator} is substituted in \Eq{pi11}, there are
three types of terms. The term that contains two factors of $\eta_e$
contributes only to the imaginary part of the self-energy and,
since we restrict ourselves here to the real part, we do not consider
it further. The remaining terms then yield
\beq
\label{pimdef}
\mbox{Re}\,\pi^{(11)}_{\mu\nu} = \pi^{(0)}_{\mu\nu} + \pi^{(m)}_{\mu\nu}\,,
\eeq
where $\pi^{(0)}_{\mu\nu}$ is the standard vacuum polarization term, while
the background dependent contribution is given by
\beq
\label{pim}
\pi^{(m)}_{\mu\nu} = -4e^2\int\frac{d^2{\vec p_\perp}}{(2\pi)^2 2E}
(f_e + f_{\overline e})
\left[\frac{L_{\mu\nu}}{k^2_\perp + 2p_\perp\cdot k_\perp} +
(k_\perp \rightarrow -k_\perp)\right]\,.
\eeq
In this formula,
\beq
\label{Lmunu}
L_{\mu\nu} = 2p_{\perp\mu} p_{\perp\nu} + p_{\perp\mu} k_{\perp\nu}
+ k_{\perp\mu} p_{\perp\nu} - g_{\perp\mu\nu}p_\perp\cdot k_\perp \,,
\eeq
\beq
\label{pE}
p^\mu_\perp = (E,\vec p_\perp)\,, \quad E = 
\sqrt{{\vec p_\perp}^{\,2} + m_e^2} \,,
\eeq
and $f_{e,\overline e}$ denote the particle and antiparticle number density
distributions given by
\beq
\label{fe}
f_{e,\overline e} = \frac{1}{e^{\beta(E \mp \mu_e)} + 1}
\eeq
with the minus(plus) sign holding for the electrons(positrons), 
respectively.
The integral in \Eq{pim} is to be interpreted in the sense of its
principal value part.

It is easily verified that $\pi^{(m)}_{\mu\nu}$ satisfies the transversality
conditions \Eqs{transversalitycond}{transversalitycond2}
and therefore it can be decomposed in the form
\beq
\pi^{(m)}_{\mu\nu} = \pi^{(m)}_T T_{\mu\nu} + \pi^{(m)}_L Q_{\mu\nu} \,,
\eeq
as we already indicated. The functions $\pi^{(m)}_{T,L}$ can be found
by projecting \Eq{pim} with the tensors $T_{\mu\nu}$ and $Q_{\mu\nu}$,
and this procedure yields
\beqa
\label{piTLm}
\pi^{(m)}_T & = & -4e^2\left(A_{\perp e} +
\frac{k^2_\perp}{\kappa^2}B_{\perp e}\right)\,,
\nonumber\\
\pi^{(m)}_L & = & 4e^2\frac{k^2_\perp}{\kappa^2}B_{\perp e}\,,
\eeqa
where
\beqa
\label{AB}
A_{\perp e} & = & \int\frac{d^2\vec p_\perp}{(2\pi)^2 2E}
(f_{e} + f_{\overline e})\left[
\frac{2m^2_e - p_\perp\cdot k_\perp}
{k^2_\perp + 2p_\perp\cdot k_\perp} + (k_\perp\rightarrow -k_\perp)\right]\,,
\nonumber\\
B_{\perp e} & = &
\int\frac{d^2\vec p_\perp}{(2\pi)^2 2E}(f_{e} + f_{\overline e})
\left[
\frac{2(p_\perp\cdot u)^2 + 2(p_\perp\cdot u)(k_\perp\cdot u) -
p_\perp\cdot k_\perp}
{k^2_\perp + 2p_\perp\cdot k_\perp} + (k_\perp\rightarrow -k_\perp)\right]
\eeqa

These two integrals are very similar to the corresponding ones
that appear in the three-dimensional case, which have been
analyzed in the literature in considerable detail,
for various limiting values of the
photon momentum and several conditions of the charged particle gas.
The methods employed there are applicable here as well.
While their evaluation is not possible for the general
form of the distribution functions, some useful results
can be obtained by considering special cases.

\subsection{Low momentum limit}

In many situations of interest, the photon momentum is such that
\beq
\label{smallkcond}
\omega, \kappa \ll E_e\,,
\eeq
where $E_e$ is the typical energy of the electrons
in the gas. In this case, borrowing the method used in
\Ref{dolivo-nieves:nucpi}, we obtain in this case
\beqa
\label{ABksmall}
B_{\perp e} & = & -\frac{1}{2}\int\frac{d^2{\vec p_\perp}}{(2\pi)^2}
\left(\frac{\vec v_{p}\cdot\vec{\kappa}}{\omega - \vec v_{p}
\cdot\vec{\kappa}}\right)\frac{d}{d{E}}(f_e + f_{\overline e})\,,\nonumber\\
A_{\perp e} & = & B_{\perp e} +
\frac{\omega}{2}\int\frac{d^2{\vec p_\perp}}{(2\pi)^2}
\left(\frac{v_{p}^2}
{\omega - \vec v_{p}\cdot\vec{\kappa}}\right)
\frac{d}{d{E}}(f_e + f_{\overline e})\,,
\eeqa
with $\vec v_{p} = \vec{p}_\perp/E$ denoting the velocity
of the particles in the background.
The above formula for $B_{\perp e}$ can be rewritten by multiplying
the integrand by the factor
\beq
\frac{1}{\omega}(\omega - \vec v_{p}\cdot\vec{\kappa} +
\vec v_{p}\cdot\vec{\kappa}) \,.
\eeq
The first two terms integrate to zero, while the third one leads to
\beq
\label{Bksmall2}
B_{\perp e} = -\frac{1}{2\omega}\int\frac{d^2{\vec p_\perp}}{(2\pi)^2}
\left(\frac{(\vec v_{p}\cdot\vec{\kappa})^2}{\omega - \vec v_{p}
\cdot\vec{\kappa}}\right)\frac{d}{d{E}}(f_e + f_{\overline e}) \,.
\eeq

We stress that the expressions given in \Eq{ABksmall} are derived from \Eq{AB}
by expanding the integrands in terms of $k/{E}$
and retaining only the terms that are dominant when
$k/{E} \rightarrow 0$. For a non-relativistic gas,
Eq.\ (\ref{ABksmall}) holds for $\omega,{\kappa} \ll m_e$.  If the
gas is extremely relativistic, Eq.\ (\ref{ABksmall}) also holds
for $\omega,{\kappa} > m_e$, subject to Eq.\ (\ref{smallkcond}).

Furthermore, up to this point no assumption has
been made regarding the nature of the electron background.
Apart from the restriction on $k$, 
\Eqs{ABksmall}{Bksmall2} hold for a relativistic
or non-relativistic gas, whether it is degenerate or not.
Accordingly, they serve as
a convenient starting point to find the dispersion relations provided that
Eq.\ (\ref{smallkcond}) is verified. As an example
we consider one specific situation below.

\subsection{Longitudinal dispersion relation}

\begin{figure}
\begin{center}
\setlength{\unitlength}{0.240900pt}
\ifx\plotpoint\undefined\newsavebox{\plotpoint}\fi
\sbox{\plotpoint}{\rule[-0.200pt]{0.400pt}{0.400pt}}%
\begin{picture}(1500,900)(0,0)
\sbox{\plotpoint}{\rule[-0.200pt]{0.400pt}{0.400pt}}%
\put(221.0,131.0){\rule[-0.200pt]{4.818pt}{0.400pt}}
\put(201,131){\makebox(0,0)[r]{ 0}}
\put(1439.0,131.0){\rule[-0.200pt]{4.818pt}{0.400pt}}
\put(221.0,222.0){\rule[-0.200pt]{4.818pt}{0.400pt}}
\put(201,222){\makebox(0,0)[r]{ 0.5}}
\put(1439.0,222.0){\rule[-0.200pt]{4.818pt}{0.400pt}}
\put(221.0,313.0){\rule[-0.200pt]{4.818pt}{0.400pt}}
\put(201,313){\makebox(0,0)[r]{ 1}}
\put(1439.0,313.0){\rule[-0.200pt]{4.818pt}{0.400pt}}
\put(221.0,404.0){\rule[-0.200pt]{4.818pt}{0.400pt}}
\put(201,404){\makebox(0,0)[r]{ 1.5}}
\put(1439.0,404.0){\rule[-0.200pt]{4.818pt}{0.400pt}}
\put(221.0,496.0){\rule[-0.200pt]{4.818pt}{0.400pt}}
\put(201,496){\makebox(0,0)[r]{ 2}}
\put(1439.0,496.0){\rule[-0.200pt]{4.818pt}{0.400pt}}
\put(221.0,587.0){\rule[-0.200pt]{4.818pt}{0.400pt}}
\put(201,587){\makebox(0,0)[r]{ 2.5}}
\put(1439.0,587.0){\rule[-0.200pt]{4.818pt}{0.400pt}}
\put(221.0,678.0){\rule[-0.200pt]{4.818pt}{0.400pt}}
\put(201,678){\makebox(0,0)[r]{ 3}}
\put(1439.0,678.0){\rule[-0.200pt]{4.818pt}{0.400pt}}
\put(221.0,769.0){\rule[-0.200pt]{4.818pt}{0.400pt}}
\put(201,769){\makebox(0,0)[r]{ 3.5}}
\put(1439.0,769.0){\rule[-0.200pt]{4.818pt}{0.400pt}}
\put(221.0,860.0){\rule[-0.200pt]{4.818pt}{0.400pt}}
\put(201,860){\makebox(0,0)[r]{ 4}}
\put(1439.0,860.0){\rule[-0.200pt]{4.818pt}{0.400pt}}
\put(221.0,131.0){\rule[-0.200pt]{0.400pt}{4.818pt}}
\put(221,90){\makebox(0,0){ 0}}
\put(221.0,840.0){\rule[-0.200pt]{0.400pt}{4.818pt}}
\put(427.0,131.0){\rule[-0.200pt]{0.400pt}{4.818pt}}
\put(427,90){\makebox(0,0){ 0.5}}
\put(427.0,840.0){\rule[-0.200pt]{0.400pt}{4.818pt}}
\put(634.0,131.0){\rule[-0.200pt]{0.400pt}{4.818pt}}
\put(634,90){\makebox(0,0){ 1}}
\put(634.0,840.0){\rule[-0.200pt]{0.400pt}{4.818pt}}
\put(840.0,131.0){\rule[-0.200pt]{0.400pt}{4.818pt}}
\put(840,90){\makebox(0,0){ 1.5}}
\put(840.0,840.0){\rule[-0.200pt]{0.400pt}{4.818pt}}
\put(1046.0,131.0){\rule[-0.200pt]{0.400pt}{4.818pt}}
\put(1046,90){\makebox(0,0){ 2}}
\put(1046.0,840.0){\rule[-0.200pt]{0.400pt}{4.818pt}}
\put(1253.0,131.0){\rule[-0.200pt]{0.400pt}{4.818pt}}
\put(1253,90){\makebox(0,0){ 2.5}}
\put(1253.0,840.0){\rule[-0.200pt]{0.400pt}{4.818pt}}
\put(1459.0,131.0){\rule[-0.200pt]{0.400pt}{4.818pt}}
\put(1459,90){\makebox(0,0){ 3}}
\put(1459.0,840.0){\rule[-0.200pt]{0.400pt}{4.818pt}}
\put(221.0,131.0){\rule[-0.200pt]{0.400pt}{175.616pt}}
\put(221.0,131.0){\rule[-0.200pt]{298.234pt}{0.400pt}}
\put(1459.0,131.0){\rule[-0.200pt]{0.400pt}{175.616pt}}
\put(221.0,860.0){\rule[-0.200pt]{298.234pt}{0.400pt}}
\put(80,495){\makebox(0,0){$\omega/\alpha$}}
\put(840,29){\makebox(0,0){$\kappa/\alpha$}}
\put(1299,820){\makebox(0,0)[r]{$\omega = \omega_\kappa$}}
\put(1319.0,820.0){\rule[-0.200pt]{24.090pt}{0.400pt}}
\put(221,131){\usebox{\plotpoint}}
\multiput(221.00,131.59)(1.123,0.482){9}{\rule{0.967pt}{0.116pt}}
\multiput(221.00,130.17)(10.994,6.000){2}{\rule{0.483pt}{0.400pt}}
\multiput(234.00,137.59)(1.267,0.477){7}{\rule{1.060pt}{0.115pt}}
\multiput(234.00,136.17)(9.800,5.000){2}{\rule{0.530pt}{0.400pt}}
\multiput(246.00,142.59)(1.123,0.482){9}{\rule{0.967pt}{0.116pt}}
\multiput(246.00,141.17)(10.994,6.000){2}{\rule{0.483pt}{0.400pt}}
\multiput(259.00,148.59)(1.267,0.477){7}{\rule{1.060pt}{0.115pt}}
\multiput(259.00,147.17)(9.800,5.000){2}{\rule{0.530pt}{0.400pt}}
\multiput(271.00,153.59)(1.378,0.477){7}{\rule{1.140pt}{0.115pt}}
\multiput(271.00,152.17)(10.634,5.000){2}{\rule{0.570pt}{0.400pt}}
\multiput(284.00,158.59)(1.033,0.482){9}{\rule{0.900pt}{0.116pt}}
\multiput(284.00,157.17)(10.132,6.000){2}{\rule{0.450pt}{0.400pt}}
\multiput(296.00,164.59)(1.378,0.477){7}{\rule{1.140pt}{0.115pt}}
\multiput(296.00,163.17)(10.634,5.000){2}{\rule{0.570pt}{0.400pt}}
\multiput(309.00,169.59)(1.267,0.477){7}{\rule{1.060pt}{0.115pt}}
\multiput(309.00,168.17)(9.800,5.000){2}{\rule{0.530pt}{0.400pt}}
\multiput(321.00,174.59)(1.378,0.477){7}{\rule{1.140pt}{0.115pt}}
\multiput(321.00,173.17)(10.634,5.000){2}{\rule{0.570pt}{0.400pt}}
\multiput(334.00,179.59)(1.267,0.477){7}{\rule{1.060pt}{0.115pt}}
\multiput(334.00,178.17)(9.800,5.000){2}{\rule{0.530pt}{0.400pt}}
\multiput(346.00,184.59)(1.378,0.477){7}{\rule{1.140pt}{0.115pt}}
\multiput(346.00,183.17)(10.634,5.000){2}{\rule{0.570pt}{0.400pt}}
\multiput(359.00,189.59)(1.267,0.477){7}{\rule{1.060pt}{0.115pt}}
\multiput(359.00,188.17)(9.800,5.000){2}{\rule{0.530pt}{0.400pt}}
\multiput(371.00,194.60)(1.797,0.468){5}{\rule{1.400pt}{0.113pt}}
\multiput(371.00,193.17)(10.094,4.000){2}{\rule{0.700pt}{0.400pt}}
\multiput(384.00,198.59)(1.267,0.477){7}{\rule{1.060pt}{0.115pt}}
\multiput(384.00,197.17)(9.800,5.000){2}{\rule{0.530pt}{0.400pt}}
\multiput(396.00,203.60)(1.797,0.468){5}{\rule{1.400pt}{0.113pt}}
\multiput(396.00,202.17)(10.094,4.000){2}{\rule{0.700pt}{0.400pt}}
\multiput(409.00,207.59)(1.267,0.477){7}{\rule{1.060pt}{0.115pt}}
\multiput(409.00,206.17)(9.800,5.000){2}{\rule{0.530pt}{0.400pt}}
\multiput(421.00,212.60)(1.797,0.468){5}{\rule{1.400pt}{0.113pt}}
\multiput(421.00,211.17)(10.094,4.000){2}{\rule{0.700pt}{0.400pt}}
\multiput(434.00,216.60)(1.651,0.468){5}{\rule{1.300pt}{0.113pt}}
\multiput(434.00,215.17)(9.302,4.000){2}{\rule{0.650pt}{0.400pt}}
\multiput(446.00,220.60)(1.797,0.468){5}{\rule{1.400pt}{0.113pt}}
\multiput(446.00,219.17)(10.094,4.000){2}{\rule{0.700pt}{0.400pt}}
\multiput(459.00,224.60)(1.651,0.468){5}{\rule{1.300pt}{0.113pt}}
\multiput(459.00,223.17)(9.302,4.000){2}{\rule{0.650pt}{0.400pt}}
\multiput(471.00,228.60)(1.797,0.468){5}{\rule{1.400pt}{0.113pt}}
\multiput(471.00,227.17)(10.094,4.000){2}{\rule{0.700pt}{0.400pt}}
\multiput(484.00,232.60)(1.651,0.468){5}{\rule{1.300pt}{0.113pt}}
\multiput(484.00,231.17)(9.302,4.000){2}{\rule{0.650pt}{0.400pt}}
\multiput(496.00,236.60)(1.797,0.468){5}{\rule{1.400pt}{0.113pt}}
\multiput(496.00,235.17)(10.094,4.000){2}{\rule{0.700pt}{0.400pt}}
\multiput(509.00,240.60)(1.651,0.468){5}{\rule{1.300pt}{0.113pt}}
\multiput(509.00,239.17)(9.302,4.000){2}{\rule{0.650pt}{0.400pt}}
\multiput(521.00,244.61)(2.695,0.447){3}{\rule{1.833pt}{0.108pt}}
\multiput(521.00,243.17)(9.195,3.000){2}{\rule{0.917pt}{0.400pt}}
\multiput(534.00,247.60)(1.651,0.468){5}{\rule{1.300pt}{0.113pt}}
\multiput(534.00,246.17)(9.302,4.000){2}{\rule{0.650pt}{0.400pt}}
\multiput(546.00,251.61)(2.695,0.447){3}{\rule{1.833pt}{0.108pt}}
\multiput(546.00,250.17)(9.195,3.000){2}{\rule{0.917pt}{0.400pt}}
\multiput(559.00,254.60)(1.651,0.468){5}{\rule{1.300pt}{0.113pt}}
\multiput(559.00,253.17)(9.302,4.000){2}{\rule{0.650pt}{0.400pt}}
\multiput(571.00,258.61)(2.695,0.447){3}{\rule{1.833pt}{0.108pt}}
\multiput(571.00,257.17)(9.195,3.000){2}{\rule{0.917pt}{0.400pt}}
\multiput(584.00,261.60)(1.651,0.468){5}{\rule{1.300pt}{0.113pt}}
\multiput(584.00,260.17)(9.302,4.000){2}{\rule{0.650pt}{0.400pt}}
\multiput(596.00,265.61)(2.695,0.447){3}{\rule{1.833pt}{0.108pt}}
\multiput(596.00,264.17)(9.195,3.000){2}{\rule{0.917pt}{0.400pt}}
\multiput(609.00,268.61)(2.472,0.447){3}{\rule{1.700pt}{0.108pt}}
\multiput(609.00,267.17)(8.472,3.000){2}{\rule{0.850pt}{0.400pt}}
\multiput(621.00,271.61)(2.695,0.447){3}{\rule{1.833pt}{0.108pt}}
\multiput(621.00,270.17)(9.195,3.000){2}{\rule{0.917pt}{0.400pt}}
\multiput(634.00,274.61)(2.472,0.447){3}{\rule{1.700pt}{0.108pt}}
\multiput(634.00,273.17)(8.472,3.000){2}{\rule{0.850pt}{0.400pt}}
\multiput(646.00,277.61)(2.695,0.447){3}{\rule{1.833pt}{0.108pt}}
\multiput(646.00,276.17)(9.195,3.000){2}{\rule{0.917pt}{0.400pt}}
\multiput(659.00,280.61)(2.472,0.447){3}{\rule{1.700pt}{0.108pt}}
\multiput(659.00,279.17)(8.472,3.000){2}{\rule{0.850pt}{0.400pt}}
\multiput(671.00,283.61)(2.695,0.447){3}{\rule{1.833pt}{0.108pt}}
\multiput(671.00,282.17)(9.195,3.000){2}{\rule{0.917pt}{0.400pt}}
\multiput(684.00,286.61)(2.472,0.447){3}{\rule{1.700pt}{0.108pt}}
\multiput(684.00,285.17)(8.472,3.000){2}{\rule{0.850pt}{0.400pt}}
\multiput(696.00,289.61)(2.695,0.447){3}{\rule{1.833pt}{0.108pt}}
\multiput(696.00,288.17)(9.195,3.000){2}{\rule{0.917pt}{0.400pt}}
\multiput(709.00,292.61)(2.472,0.447){3}{\rule{1.700pt}{0.108pt}}
\multiput(709.00,291.17)(8.472,3.000){2}{\rule{0.850pt}{0.400pt}}
\multiput(721.00,295.61)(2.695,0.447){3}{\rule{1.833pt}{0.108pt}}
\multiput(721.00,294.17)(9.195,3.000){2}{\rule{0.917pt}{0.400pt}}
\multiput(734.00,298.61)(2.472,0.447){3}{\rule{1.700pt}{0.108pt}}
\multiput(734.00,297.17)(8.472,3.000){2}{\rule{0.850pt}{0.400pt}}
\put(746,301.17){\rule{2.700pt}{0.400pt}}
\multiput(746.00,300.17)(7.396,2.000){2}{\rule{1.350pt}{0.400pt}}
\multiput(759.00,303.61)(2.472,0.447){3}{\rule{1.700pt}{0.108pt}}
\multiput(759.00,302.17)(8.472,3.000){2}{\rule{0.850pt}{0.400pt}}
\multiput(771.00,306.61)(2.695,0.447){3}{\rule{1.833pt}{0.108pt}}
\multiput(771.00,305.17)(9.195,3.000){2}{\rule{0.917pt}{0.400pt}}
\multiput(784.00,309.61)(2.472,0.447){3}{\rule{1.700pt}{0.108pt}}
\multiput(784.00,308.17)(8.472,3.000){2}{\rule{0.850pt}{0.400pt}}
\put(796,312.17){\rule{2.700pt}{0.400pt}}
\multiput(796.00,311.17)(7.396,2.000){2}{\rule{1.350pt}{0.400pt}}
\multiput(809.00,314.61)(2.472,0.447){3}{\rule{1.700pt}{0.108pt}}
\multiput(809.00,313.17)(8.472,3.000){2}{\rule{0.850pt}{0.400pt}}
\put(821,317.17){\rule{2.700pt}{0.400pt}}
\multiput(821.00,316.17)(7.396,2.000){2}{\rule{1.350pt}{0.400pt}}
\multiput(834.00,319.61)(2.472,0.447){3}{\rule{1.700pt}{0.108pt}}
\multiput(834.00,318.17)(8.472,3.000){2}{\rule{0.850pt}{0.400pt}}
\put(846,322.17){\rule{2.700pt}{0.400pt}}
\multiput(846.00,321.17)(7.396,2.000){2}{\rule{1.350pt}{0.400pt}}
\multiput(859.00,324.61)(2.472,0.447){3}{\rule{1.700pt}{0.108pt}}
\multiput(859.00,323.17)(8.472,3.000){2}{\rule{0.850pt}{0.400pt}}
\put(871,327.17){\rule{2.700pt}{0.400pt}}
\multiput(871.00,326.17)(7.396,2.000){2}{\rule{1.350pt}{0.400pt}}
\multiput(884.00,329.61)(2.472,0.447){3}{\rule{1.700pt}{0.108pt}}
\multiput(884.00,328.17)(8.472,3.000){2}{\rule{0.850pt}{0.400pt}}
\put(896,332.17){\rule{2.700pt}{0.400pt}}
\multiput(896.00,331.17)(7.396,2.000){2}{\rule{1.350pt}{0.400pt}}
\put(909,334.17){\rule{2.500pt}{0.400pt}}
\multiput(909.00,333.17)(6.811,2.000){2}{\rule{1.250pt}{0.400pt}}
\multiput(921.00,336.61)(2.695,0.447){3}{\rule{1.833pt}{0.108pt}}
\multiput(921.00,335.17)(9.195,3.000){2}{\rule{0.917pt}{0.400pt}}
\put(934,339.17){\rule{2.500pt}{0.400pt}}
\multiput(934.00,338.17)(6.811,2.000){2}{\rule{1.250pt}{0.400pt}}
\put(946,341.17){\rule{2.700pt}{0.400pt}}
\multiput(946.00,340.17)(7.396,2.000){2}{\rule{1.350pt}{0.400pt}}
\multiput(959.00,343.61)(2.472,0.447){3}{\rule{1.700pt}{0.108pt}}
\multiput(959.00,342.17)(8.472,3.000){2}{\rule{0.850pt}{0.400pt}}
\put(971,346.17){\rule{2.700pt}{0.400pt}}
\multiput(971.00,345.17)(7.396,2.000){2}{\rule{1.350pt}{0.400pt}}
\put(984,348.17){\rule{2.500pt}{0.400pt}}
\multiput(984.00,347.17)(6.811,2.000){2}{\rule{1.250pt}{0.400pt}}
\put(996,350.17){\rule{2.700pt}{0.400pt}}
\multiput(996.00,349.17)(7.396,2.000){2}{\rule{1.350pt}{0.400pt}}
\put(1009,352.17){\rule{2.500pt}{0.400pt}}
\multiput(1009.00,351.17)(6.811,2.000){2}{\rule{1.250pt}{0.400pt}}
\multiput(1021.00,354.61)(2.695,0.447){3}{\rule{1.833pt}{0.108pt}}
\multiput(1021.00,353.17)(9.195,3.000){2}{\rule{0.917pt}{0.400pt}}
\put(1034,357.17){\rule{2.500pt}{0.400pt}}
\multiput(1034.00,356.17)(6.811,2.000){2}{\rule{1.250pt}{0.400pt}}
\put(1046,359.17){\rule{2.700pt}{0.400pt}}
\multiput(1046.00,358.17)(7.396,2.000){2}{\rule{1.350pt}{0.400pt}}
\put(1059,361.17){\rule{2.500pt}{0.400pt}}
\multiput(1059.00,360.17)(6.811,2.000){2}{\rule{1.250pt}{0.400pt}}
\put(1071,363.17){\rule{2.700pt}{0.400pt}}
\multiput(1071.00,362.17)(7.396,2.000){2}{\rule{1.350pt}{0.400pt}}
\put(1084,365.17){\rule{2.500pt}{0.400pt}}
\multiput(1084.00,364.17)(6.811,2.000){2}{\rule{1.250pt}{0.400pt}}
\put(1096,367.17){\rule{2.700pt}{0.400pt}}
\multiput(1096.00,366.17)(7.396,2.000){2}{\rule{1.350pt}{0.400pt}}
\put(1109,369.17){\rule{2.500pt}{0.400pt}}
\multiput(1109.00,368.17)(6.811,2.000){2}{\rule{1.250pt}{0.400pt}}
\put(1121,371.17){\rule{2.700pt}{0.400pt}}
\multiput(1121.00,370.17)(7.396,2.000){2}{\rule{1.350pt}{0.400pt}}
\put(1134,373.17){\rule{2.500pt}{0.400pt}}
\multiput(1134.00,372.17)(6.811,2.000){2}{\rule{1.250pt}{0.400pt}}
\put(1146,375.17){\rule{2.700pt}{0.400pt}}
\multiput(1146.00,374.17)(7.396,2.000){2}{\rule{1.350pt}{0.400pt}}
\put(1159,377.17){\rule{2.500pt}{0.400pt}}
\multiput(1159.00,376.17)(6.811,2.000){2}{\rule{1.250pt}{0.400pt}}
\put(1171,379.17){\rule{2.700pt}{0.400pt}}
\multiput(1171.00,378.17)(7.396,2.000){2}{\rule{1.350pt}{0.400pt}}
\put(1184,381.17){\rule{2.500pt}{0.400pt}}
\multiput(1184.00,380.17)(6.811,2.000){2}{\rule{1.250pt}{0.400pt}}
\put(1196,383.17){\rule{2.700pt}{0.400pt}}
\multiput(1196.00,382.17)(7.396,2.000){2}{\rule{1.350pt}{0.400pt}}
\put(1209,385.17){\rule{2.500pt}{0.400pt}}
\multiput(1209.00,384.17)(6.811,2.000){2}{\rule{1.250pt}{0.400pt}}
\put(1221,387.17){\rule{2.700pt}{0.400pt}}
\multiput(1221.00,386.17)(7.396,2.000){2}{\rule{1.350pt}{0.400pt}}
\put(1234,389.17){\rule{2.500pt}{0.400pt}}
\multiput(1234.00,388.17)(6.811,2.000){2}{\rule{1.250pt}{0.400pt}}
\put(1246,391.17){\rule{2.700pt}{0.400pt}}
\multiput(1246.00,390.17)(7.396,2.000){2}{\rule{1.350pt}{0.400pt}}
\put(1259,393.17){\rule{2.500pt}{0.400pt}}
\multiput(1259.00,392.17)(6.811,2.000){2}{\rule{1.250pt}{0.400pt}}
\put(1271,395.17){\rule{2.700pt}{0.400pt}}
\multiput(1271.00,394.17)(7.396,2.000){2}{\rule{1.350pt}{0.400pt}}
\put(1284,396.67){\rule{2.891pt}{0.400pt}}
\multiput(1284.00,396.17)(6.000,1.000){2}{\rule{1.445pt}{0.400pt}}
\put(1296,398.17){\rule{2.700pt}{0.400pt}}
\multiput(1296.00,397.17)(7.396,2.000){2}{\rule{1.350pt}{0.400pt}}
\put(1309,400.17){\rule{2.500pt}{0.400pt}}
\multiput(1309.00,399.17)(6.811,2.000){2}{\rule{1.250pt}{0.400pt}}
\put(1321,402.17){\rule{2.700pt}{0.400pt}}
\multiput(1321.00,401.17)(7.396,2.000){2}{\rule{1.350pt}{0.400pt}}
\put(1334,404.17){\rule{2.500pt}{0.400pt}}
\multiput(1334.00,403.17)(6.811,2.000){2}{\rule{1.250pt}{0.400pt}}
\put(1346,406.17){\rule{2.700pt}{0.400pt}}
\multiput(1346.00,405.17)(7.396,2.000){2}{\rule{1.350pt}{0.400pt}}
\put(1359,407.67){\rule{2.891pt}{0.400pt}}
\multiput(1359.00,407.17)(6.000,1.000){2}{\rule{1.445pt}{0.400pt}}
\put(1371,409.17){\rule{2.700pt}{0.400pt}}
\multiput(1371.00,408.17)(7.396,2.000){2}{\rule{1.350pt}{0.400pt}}
\put(1384,411.17){\rule{2.500pt}{0.400pt}}
\multiput(1384.00,410.17)(6.811,2.000){2}{\rule{1.250pt}{0.400pt}}
\put(1396,413.17){\rule{2.700pt}{0.400pt}}
\multiput(1396.00,412.17)(7.396,2.000){2}{\rule{1.350pt}{0.400pt}}
\put(1409,414.67){\rule{2.891pt}{0.400pt}}
\multiput(1409.00,414.17)(6.000,1.000){2}{\rule{1.445pt}{0.400pt}}
\put(1421,416.17){\rule{2.700pt}{0.400pt}}
\multiput(1421.00,415.17)(7.396,2.000){2}{\rule{1.350pt}{0.400pt}}
\put(1434,418.17){\rule{2.500pt}{0.400pt}}
\multiput(1434.00,417.17)(6.811,2.000){2}{\rule{1.250pt}{0.400pt}}
\put(1446,420.17){\rule{2.700pt}{0.400pt}}
\multiput(1446.00,419.17)(7.396,2.000){2}{\rule{1.350pt}{0.400pt}}
\put(1299,779){\makebox(0,0)[r]{$\omega = \kappa$}}
\multiput(1319,779)(20.756,0.000){5}{\usebox{\plotpoint}}
\put(1419,779){\usebox{\plotpoint}}
\put(221,131){\usebox{\plotpoint}}
\put(221.00,131.00){\usebox{\plotpoint}}
\put(239.94,139.48){\usebox{\plotpoint}}
\put(258.89,147.95){\usebox{\plotpoint}}
\put(277.93,156.20){\usebox{\plotpoint}}
\put(296.97,164.45){\usebox{\plotpoint}}
\put(315.93,172.89){\usebox{\plotpoint}}
\put(334.87,181.36){\usebox{\plotpoint}}
\put(353.90,189.65){\usebox{\plotpoint}}
\put(372.94,197.90){\usebox{\plotpoint}}
\put(391.92,206.30){\usebox{\plotpoint}}
\put(410.86,214.77){\usebox{\plotpoint}}
\put(429.87,223.09){\usebox{\plotpoint}}
\put(448.91,231.34){\usebox{\plotpoint}}
\put(467.90,239.71){\usebox{\plotpoint}}
\put(486.76,248.38){\usebox{\plotpoint}}
\put(505.73,256.74){\usebox{\plotpoint}}
\put(524.58,265.38){\usebox{\plotpoint}}
\put(543.53,273.77){\usebox{\plotpoint}}
\put(562.64,281.82){\usebox{\plotpoint}}
\put(581.65,290.10){\usebox{\plotpoint}}
\put(600.50,298.73){\usebox{\plotpoint}}
\put(619.42,307.21){\usebox{\plotpoint}}
\put(638.52,315.26){\usebox{\plotpoint}}
\put(657.57,323.45){\usebox{\plotpoint}}
\put(676.42,332.08){\usebox{\plotpoint}}
\put(695.30,340.65){\usebox{\plotpoint}}
\put(714.41,348.70){\usebox{\plotpoint}}
\put(733.49,356.80){\usebox{\plotpoint}}
\put(752.34,365.44){\usebox{\plotpoint}}
\put(771.19,374.09){\usebox{\plotpoint}}
\put(790.13,382.55){\usebox{\plotpoint}}
\put(809.07,391.03){\usebox{\plotpoint}}
\put(828.11,399.28){\usebox{\plotpoint}}
\put(847.16,407.53){\usebox{\plotpoint}}
\put(866.12,415.97){\usebox{\plotpoint}}
\put(885.06,424.44){\usebox{\plotpoint}}
\put(904.08,432.73){\usebox{\plotpoint}}
\put(923.13,440.98){\usebox{\plotpoint}}
\put(942.10,449.38){\usebox{\plotpoint}}
\put(961.05,457.85){\usebox{\plotpoint}}
\put(980.05,466.18){\usebox{\plotpoint}}
\put(999.10,474.43){\usebox{\plotpoint}}
\put(1018.09,482.79){\usebox{\plotpoint}}
\put(1036.94,491.47){\usebox{\plotpoint}}
\put(1055.92,499.81){\usebox{\plotpoint}}
\put(1074.77,508.45){\usebox{\plotpoint}}
\put(1093.72,516.86){\usebox{\plotpoint}}
\put(1112.82,524.91){\usebox{\plotpoint}}
\put(1131.84,533.17){\usebox{\plotpoint}}
\put(1150.69,541.80){\usebox{\plotpoint}}
\put(1169.60,550.30){\usebox{\plotpoint}}
\put(1188.71,558.35){\usebox{\plotpoint}}
\put(1207.76,566.52){\usebox{\plotpoint}}
\put(1226.61,575.16){\usebox{\plotpoint}}
\put(1245.48,583.74){\usebox{\plotpoint}}
\put(1264.59,591.79){\usebox{\plotpoint}}
\put(1283.68,599.88){\usebox{\plotpoint}}
\put(1302.53,608.51){\usebox{\plotpoint}}
\put(1321.37,617.17){\usebox{\plotpoint}}
\put(1340.32,625.63){\usebox{\plotpoint}}
\put(1359.26,634.11){\usebox{\plotpoint}}
\put(1378.30,642.37){\usebox{\plotpoint}}
\put(1397.34,650.62){\usebox{\plotpoint}}
\put(1416.31,659.04){\usebox{\plotpoint}}
\put(1435.25,667.52){\usebox{\plotpoint}}
\put(1454.27,675.82){\usebox{\plotpoint}}
\put(1459,678){\usebox{\plotpoint}}
\put(221.0,131.0){\rule[-0.200pt]{0.400pt}{175.616pt}}
\put(221.0,131.0){\rule[-0.200pt]{298.234pt}{0.400pt}}
\put(1459.0,131.0){\rule[-0.200pt]{0.400pt}{175.616pt}}
\put(221.0,860.0){\rule[-0.200pt]{298.234pt}{0.400pt}}
\end{picture}
\end{center}
\caption[]{
Plot of the longitudinal dispersion relation given in
Eq.\ (\ref{longdisprelsol}).
\label{fig:longdisprelexample}
}
\end{figure}
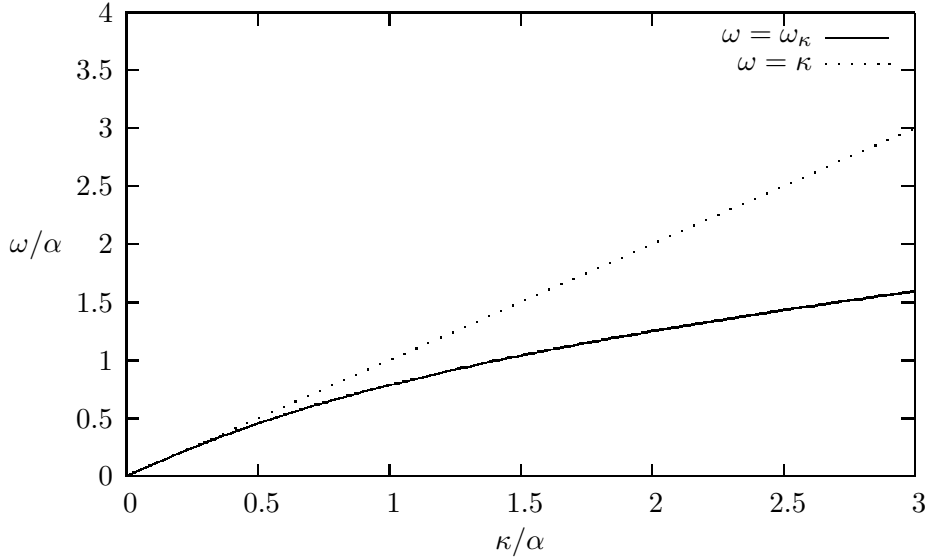

We consider the case of a non-relativistic electron gas, in the regime
in which \Eq{smallkcond} is valid, 
and seek the real solutions to the longitudinal dispersion relations
in the long wavelength limit,
\beq
\omega \gg \kappa v_e\,,
\eeq
where $v_e$ is the
typical velocity of the electrons. Then, neglecting $f_{\overline e}$
in \Eq{Bksmall2}, the calculation of $B_e$ in this limit gives
\beq
\label{Benrclassical}
B_{\perp e} = \frac{n_e\kappa^2}{4m_e\omega^2}\,,
\eeq
and using \Eqss{pieff}{pimdef}{piTLm},
\beq
\mbox{Re}\,\pi_L = \frac{e^2 n_e}{m_e\omega^2}
\left(\omega^2 - \kappa^2\right)\,.
\eeq
A real solution of \Eq{longdisprel} is obtained for $\omega < \kappa$ by
solving
\beq
\left(\frac{e^2 n_e}{2m_e}\right)\frac{\sqrt{\kappa^2 - \omega^2}}{\omega^2}
= 1\,,
\eeq
which gives
\beq
\label{longdisprelsol}
\omega^2_\kappa = \frac{1}{2}\alpha^2\left[
\left(1 + \frac{4\kappa^2}{\alpha^2}\right)^{1/2} - 1\right]\,,
\eeq
where
\beq
\alpha = \frac{e^2 n_e}{2m_e} \,.
\eeq
The function $\omega_\kappa$ is depicted in Fig. \ref{fig:longdisprelexample}.
It follows from \Eq{longdisprelsol} that $\omega_\kappa \simeq \kappa$ for
$\kappa \ll \alpha$, while for $\kappa \gg \alpha$ it is given approximately by
\beq
\label{longdisprelexample}
\omega^2_\kappa \simeq \alpha \kappa \,.
\eeq
This solution is valid as long as
\beq
\kappa < \frac{\alpha}{v^2_e} \,,
\eeq
so that the condition $\omega \gg \kappa v_e$ is satisfied
and therefore \Eq{Benrclassical} remains valid.
%

This result, which implies that in this example system the plasma frequency
is momentum-dependent, is known in the plasma physics
and condensed matter literature\cite{bret,fetter}. In those contexts,
it is derived using the
\emph{static local field correction approximation},
in which the dynamic corrections are replaced by their static
($\omega \rightarrow 0$) values\cite{ichimaru2}.
In our notation, that approximation corresponds to using the expressions
\beqa
\hat\Delta(k_\perp) & \simeq & \lim_{\omega\rightarrow 0}\, \hat\Delta(k_\perp)
= \frac{1}{2\kappa}\,,\nonumber\\
\mbox{Re}\,\pi_L & \simeq & \left. \mbox{Re}\,\pi_L\right|_{\omega \ll \kappa}
= -\,\frac{e^2 n_e \kappa^2}{m_e\omega^2}\,,
\eeqa
in \Eq{longdisprel}. The solution of the equation thus obtained is indeed
given by \Eq{longdisprelexample}. But, as we have seen,
\Eq{longdisprelexample} is an approximation to the solution given in
\Eq{longdisprelsol}, which we have obtained by using the complete expressions
for the propagator and self-energy.

Therefore, in our case, by using the complete
inverse propagator in \Eq{longdisprel}, we are able to obtain the
dispersion relation given in \Eq{longdisprelsol}, which is valid for all
values of $\kappa$, including $\kappa \simeq \alpha$
which are outside the range
of validity ($\kappa \gg \alpha$) of the \emph{static local field
correction approximation}.

Our motivation for going through this exercise here is to evidence
that some known results are not only reproduced, but also generalized,
using the method we plan to use, which in turn indicate that they can
be applied systematically to study the problems that we have mentioned
in this type of system.
%
%
\section{Conclusions}
\label{sec:conclusions}

In the present work we have formulated the TFT
approach to the model of the two-dimensional plasma layer, that is,
a system in which the electrons are confined to a plane sheet. 
As emphasized in the Introduction, this is not equivalent to what
is usually called $QED_3$ (or $QED$ in 2+1 dimensions)
which describes a two-dimensional cross section of
a system that has cylindrical symmetry and
the physics is independent of the $z$ coordinate.
Thus, for example, while the \emph{electrons} in $QED_3$ are really
lines of charge in the three-dimensional world and the Coulomb potential
between them is logarithmic, in
the system we are considering the electrons are ordinary point charges,
which are confined to the $z = 0$ plane, but the Coulomb potential between
them is the ordinary $1/r$ potential.

An important step in this direction was to determine
the appropriate set of thermal propagators. As an application,
we performed the one-loop calculation of the photon
self-energy in that medium, and we considered
in detail the calculation of the longitudinal photon dispersion relation.
We made contact with previous calculations of that quantity,
that had been obtained using other approaches,
and in particular we showed that our results reduce to those known results
when the appropriate limits are taken and/or approximations are made.
However, the formulas we obtained are more general and can be used
for a wider range of conditions in which those approximations and limits
are not justified. We considered the simplest situation of an ordinary
gas of electrons, which are confined to a plane sheet, but are otherwise
free.  However the method allows us to 
consider variations of the model in a systematic way, such as
the effects of anisotropies and/or external fields.

There is a considerable amount of literature on the applications
of Thermal Field Theory (TFT) methods to the study of
the properties of the electron gas (plasmon dispersion relations and so on)
in the three-dimensional space. Although those systems have been
studied by a variety of methods, the application of TFT
methods to study them have been useful in many ways.

We believe that the applications to the two-dimensional sheet
that we have described, are novel TFT calculations which are expected
to be equally attractive and useful. For example,
they can have astrophysical applications\cite{lyutikov,asseo,gedalin};
they establish a good point of contact between
the applications of TFT methods and dominant themes in plasma physics
(such as the effects of anisotropies and instability studies);
they may be are relevant for laboratory experiments 
in fundamental disciplines of physics such as
plasma physics\cite{jiang} and condensed matter\cite{nagao,uchida}.

Such calculations are also interesting in their own right because
they can be useful in the study of physical systems of current interest
in which a plasma is confined to a layer\cite{caldas} or a wire\cite{faccioli},
and they may also be helpful for studying and understanding analogous
effects and issues in more complicated systems such as the non-abelian plasmas.

\end{document}